\documentstyle[12pt,epsfig,rotating]{article}
 
\begin{document}

\begin{flushright}
{\bf 14 November, 1996}
\end{flushright}

\vskip 1.0 cm

\centerline{\large\bf An All-Solid State Central Tracker for the Proposed}
\centerline{\large\bf DESY Electron-Positron Linear Collider}

\vskip 2.0 cm

\centerline{B. J. King}

\vskip 0.5 cm

\centerline{DESY}
\centerline{F-OPAL}
\centerline{Notkestrasse 85}
\centerline{22607 Hamburg}
\centerline{Deutschland}
\vskip 0.5 cm
\centerline{email: KING$@$OPL01.DESY.DE}
\vskip 1.5 cm

\centerline{\large\bf Abstract}
\vskip 0.5 cm

    This report describes an all-solid state central tracker which
is intended for use in a detector at the proposed DESY 500 GeV
electron-positron linear collider or a similar accelerator.
The precise position measurements from position-sensitive silicon
detectors give the tracker an outstanding momentum resolution for
high momentum tracks:
${\rm \sigma_p/p^2 = 3.6 \times 10^{-5}\;(GeV/c)^{-1}}$
for tracks perpendicular to the beam-line.
The report concludes with an example layout for a detector
which uses this central tracker.

\pagebreak

\section{Introduction}

   Rapid progress in solid-state detector technology has made
large area solid-state central trackers an attractive option for
collider experiments. Examples are the CMS~\cite{cmstp,cmscost} and
ATLAS~\cite{atlastp,atlascost} experiments at the LHC, the latter of
which will use 42 ${\rm m}^2$ of microstrip detectors and approximately
4 ${\rm m}^2$ of pixel detectors in its central tracker.

   This report describes an all-silicon central tracker (CT)
which is intended for use at a TeV-scale electron-positron
linear collider, and which features an outstanding momentum
resolution of
$3.6 \times 10^{-5}\; ({\rm GeV/c})^{-1}$
for high momentum tracks perpendicular to the beam-line.

\begin{figure}[tb]
\begin{center}
\mbox{\epsfxsize=11.0 cm\epsffile[0 100 565 450]{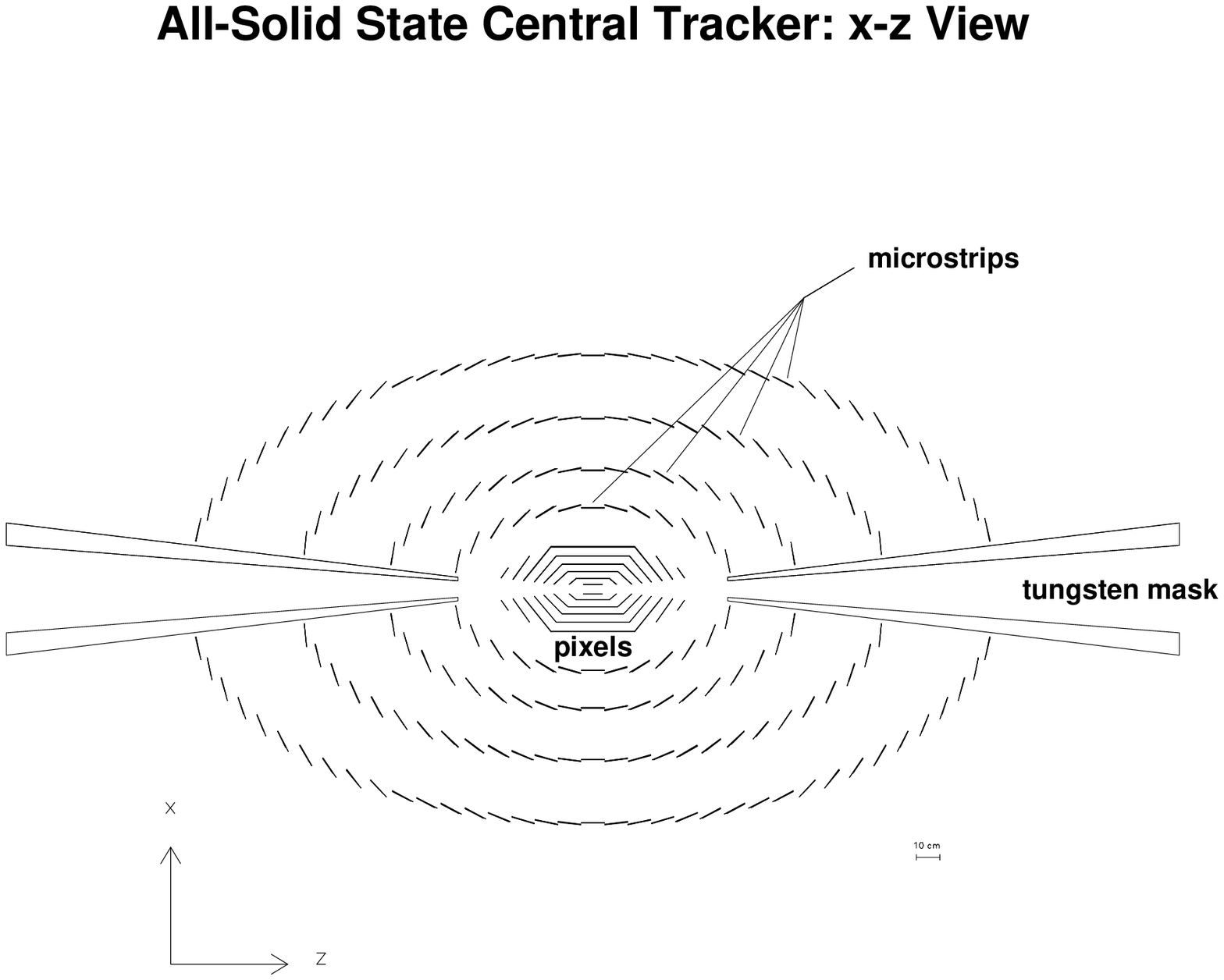}}
\end{center}
\caption{
{\small \bf
The all-silicon central tracker design. The tracker has a diameter
of 2.0 metres and a length of 3.4 metres.
The tiles in the outermost shell represent 3 layers of
microstrip detectors in U-V-X configuration. In the 3
shells near mid-radius the tiles represent back-to-back
microstrip detectors in small angle stereo configuration.
The 6 vertexing layers consist of pixel detectors. The
conical sections of the tungsten radiation masks are also
shown.}
}
\label{fig:ctxz}
\end{figure}

\begin{figure}[tb]
\begin{center}
\mbox{\epsfxsize=11.0cm\epsffile[0 60 565 500]{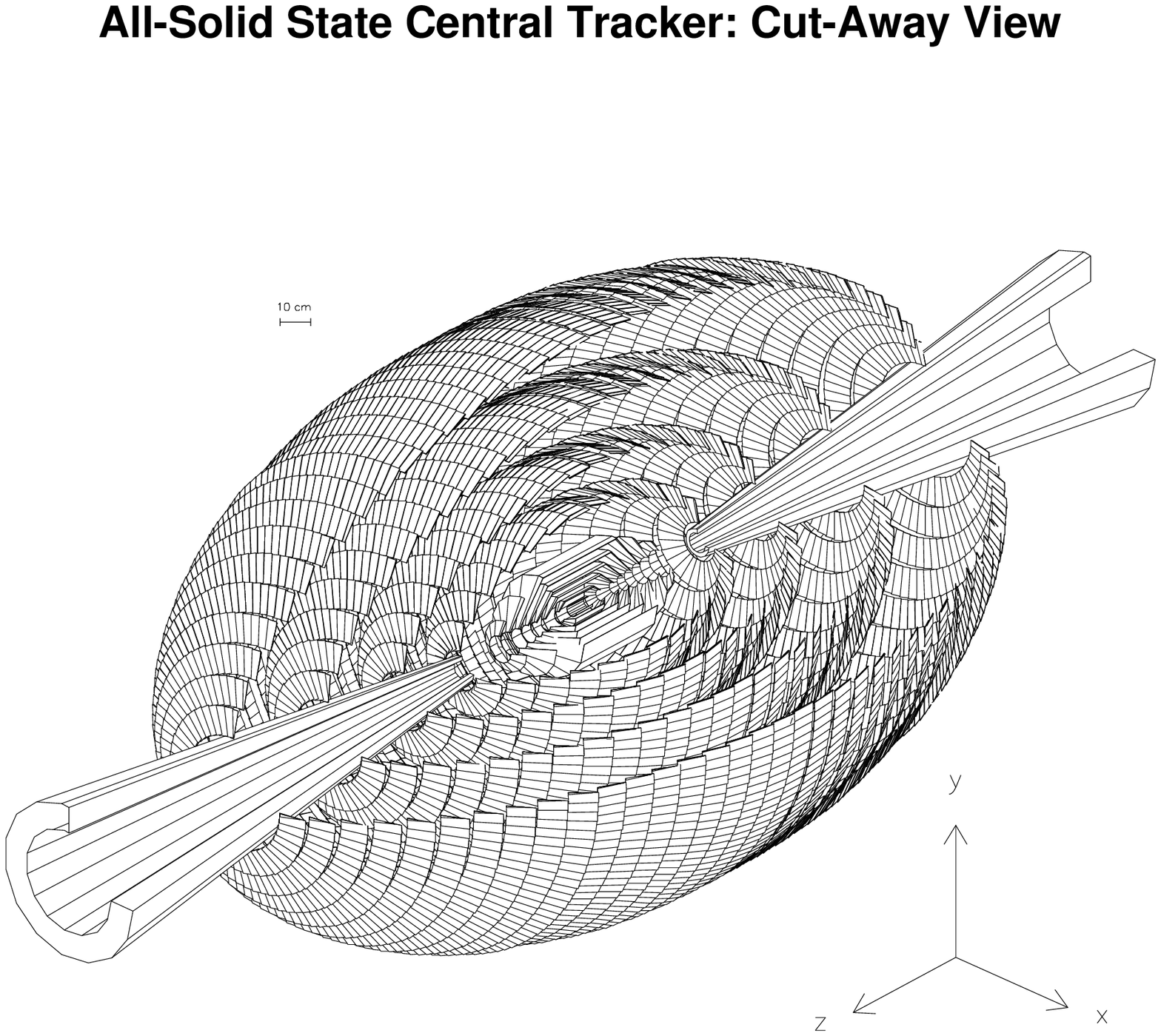}}
\end{center}
\caption{
{\small \bf
A 3-dimensional cut-away view of the central tracker and tungsten
radiation masks.
}
}
\label{fig:ctproj}
\end{figure}

   A cross-section of the CT is displayed in figure~\ref{fig:ctxz},
and figure~\ref{fig:ctproj} gives a 3-dimensional cut-away view. Silicon
position-sensitive detectors are arranged in 10 concentric ellipsoidal
shells around the interaction point (IP): pixel detectors in the
6 innermost layers and microstrip detectors in the 4 large outer
layers. The outer radius of the tracker is 1.00 m,
the half-length is 1.70 m and the CT covers 99.3 percent of
the solid angle. The minimum polar angle, 120 milliradians,
coincides with the outer edge of the tungsten radiation mask.
The ellipsoidal geometry can easily accommodate tracking coverage
down to even smaller polar angles, if it is found that the
tungsten mask can be reduced in size.

   The 6 innermost shells comprise 1.75 square metres of silicon
pixel detectors while the 3 shells at mid-radius use back-to-back
silicon microstrip detectors in stereo geometry and the
outermost shell uses a triple layer of microstrips in
``U-V-X'' geometry. The total
area of strip detectors is 96 square metres.

   Details of the
tracker layout are given in section 2 and the silicon detectors
are discussed in section 3. Sections 4, 5 and 6 address the technical
issues of pattern recognition, mechanical support structure and
mechanical services, and the alignment of the silicon wafers, respectively.
The performance of the tracker for vertexing and momentum
resolution is discussed in section 7, and the cost of the
tracker in section 8. Finally, an example design for the
entire detector is presented in section 9.

    The central tracker and detector designs have been specifically
geared towards the two linear collider options which have been
proposed for the DESY Laboratory.
In this report,
the two options will be referred to as ``TESLA'' and ``S-band'',
and the generic linear collider as the ``TLC'' (``TeV-scale
linear collider''). The central tracker design will often
be abbreviated to ``CT'' and the overall detector design
as ``TLD'' (``detector for a TeV-scale linear collider'').

\section{Layout of Central Tracker}

    The main design goals for the layout of the CT have been to
provide excellent vertexing and nearly optimal momentum resolution
for high-momentum tracks, consistent with robust pattern recognition
and convenient mechanical layout.
Of necessity, there is some level of compromise between the
competing design issues. For example, the mid-radius region of
the tracker is the most critical for determining the sagitta of
high-momentum tracks. However, material in this region degrades
the measurement of low-momentum tracks. The demands of pattern
recognition and mechanical support also favour spreading out the shells
rather than concentrating them at mid-radius.

   The use of many small solid-state detectors, as opposed to a
gas-based tracking chamber, introduces a much greater freedom in
the choice of tracker layout. The ellipsoidal shell geometry which
was chosen for the CT has several
attractive features relative to, say, a barrel-plus-endcap geometry:
\begin{itemize}
\item  all of the ellipsoidal surface is reasonably projective to
       the interaction point (IP). The silicon
       wafers can be made projective with little additional tilting.
\item  the explicit radial ordering of the shells is helpful for
       pattern recognition. (See section 4.)
\item  the regions between the shells are available for a stiff,
       light 3-dimensional mechanical support structure and also
       provide natural outlets for routing cables and cooling pipes.
       (See section 5.)
\item  tracking coverage can be provided down to small polar angles
       with no barrel-to-endcap cross-over region and with a smooth
       dependence of the resolution functions on polar angle.
\item  the smaller outer surface area of the tracker can be
       surrounded by a cheaper, more projective electromagnetic
       calorimeter than would be the case with, say, a cylindrical
       geometry. (See section 9.)
\end{itemize}
Partly because of these advantages, the ellipsoidal geometry has
already been chosen for the tracking spectrometer of the CLAS detector
at CEBAF~\cite{clas} (using curved drift chambers) and
the BaBar silicon vertex tracker~\cite{babar} at the PEP-II asymmetric
electron-positron collider at SLAC.

   Pixel detectors (either CCD's or APS's -- see the following
section) are strongly preferred over microstrips for
pattern recognition in the 6 innermost layers, where beamstrahlung
electrons will produce of order 1000 hits -- see section 4.
In contrast, the hit multiplicities in the 4 large outer shells
are dominated by the physics event itself, giving only of order
100 hits spread over areas of a square meter or more. Stereo layers
of microstrips (or a triple layer, in the case of the outermost
shell) give adequate granularity for these shells at an
affordable price.

The strip detectors are tilted to make an angle which is projective
to the interaction point (IP). Projective detectors give the
optimal point resolution for stiff tracks, and this geometry also
covers the CT acceptance using the smallest possible area of
detectors, saving in cost and material budget.

    The wafers overlap
in both the polar and azimuthal views, as viewed from the IP,
which makes the CT hermetic to tracks originating near the IP. In
fact, the CT is also hermetic to soft tracks from the IP, since their
transverse momentum is initially in the radial direction and
the track's polar angle decreases (for forward-going tracks)
as the track spirals and the radial momentum component is 
transferred into the azimuthal component. Thus, these dipping tracks
see even more overlap in the wafers than stiff tracks. The
hermeticity of the CT has been checked in Monte Carlo-based
track simulations.

   The 6 pixel layers near the interaction point (IP) are the most
important for finding
decay vertices from beauty and charm hadrons and tau leptons. For
this reason, they will sometimes be referred to as the ``vertexing
layers'' or, collectively, as the ``vertex detector''. However, it
should be remembered that, as with the other tasks of the CT,
information from the entire tracker is used for vertex finding.

    The layout of the vertexing layers is
similar to that of the BaBar silicon vertex tracker~\cite{babar}.
The innermost layer, at a radius of 2 cm, is a cylinder, but the
remaining 5 concentric layers consist of a central cylinder with
end-cap cones at each end. Two additional end-caps at each end of the
vertex detector ensure 6 layers of vertexing for all tracks down to
the minimum polar angle of the tracker acceptance, at 120 mrad from
the beam direction.

   The barrel regions of the 6 layers have radii of 2, 4, 7, 10, 13
and 16 cm. The innermost vertexing layer is placed directly outside
the beam-pipe to minimize the extrapolation
distance to decay vertices. The 2 cm radius of this layer is well
matched to the decay lengths of relativistically boosted beauty
and charm hadrons -- about 1 cm for energies of 100 GeV -- while
avoiding the rapidly increasing backgrounds at smaller radii.

   The nearly-spherical geometry of the vertex detector means that
the wafers are all fairly projective to the IP, giving nearly optimal
point resolutions for almost all tracks. This geometry is particularly
appropriate for electron-positron colliders, as opposed to e-p or p-p
colliders, since electron-positron colliders have a beam spot which
is very small compared to the scale of the detector wafers. Also, the
energies and, hence, lifetime
distributions of beauty, charm and tau particles don't depend
strongly on the polar angle.

   The CT is elongated along the direction of the magnetic field
(i.e. parallel to the beam axis) to improve the momentum
resolution for forward-going tracks. The benefits of elongating the CT
must be weighed against the disadvantages for the tracker and the
sub-detectors outside it, such as a more expensive, less
projective electromagnetic calorimeter and a longer magnet.

   Typically, trackers at electron-positron colliders are less elongated
than their counterparts at e-p and p-p colliders, where the interesting
physics processes are more concentrated at small angles to the beam direction.
This reflects the hard, point-like nature of electron-positron
collisions relative to the softer interactions of composite protons.
Following this expectation, the central region of the detector is
found to be relatively important for the physics processes that have
been studied so far for the TLC~\cite{desyphysics}.

   For the chosen
aspect ratio of the tracker, eighty-seven percent of the CT's solid
angle has a momentum resolution within a factor of two of that
perpendicular to the beam-line. This is similar to, or slightly greater
than, the corresponding percentage for today's trackers at high
energy electron-positron colliders. Even at the minimum polar angle
of the CT acceptance, at 120 mrad from the beam-line, the momentum
resolution is still sufficient to determine the
charge of tracks up to the beam energy. (See section 7 for details
on the angular dependence of the momentum distribution.)

   The stereo layers of microstrips which instrument the 4 large
tracker shells consist of tiles containing two back-to-back
single-sided silicon detectors, for reasons discussed in the next
section. For optimal resolution of tracks' bend coordinate, all of
the tiles are oriented so that the stereo strip directions are
bisected by a plane
containing the beam-line axis. In addition, the tiles in the outermost
shell incorporate a third single-sided detector whose strips are
oriented in this plane, bisecting the stereo directions. This
``U-V-X'' geometry gives the redundancy of 3 position measurements
with the full lever-arm of the tracker. 

   The main motivation for using stereo strips is that the two
hits produced by a track can immediately be combined into a space point
as the first step of a pattern recognition algorithm -- see section 4.
In this respect,
the stereo configuration also has important advantages over
back-to-back orthogonal strips, since hit ambiguities (``ghost hits'')
will usually only occur when a second track crosses the wafer within a defined
area of order (stereo angle)--times--(strip length) around the first,
which will correspond to only a small fraction of the wafer area.
In addition, the smaller overlap area reduces the chance that a random
noise hit will be added on to a track hit to contribute a ghost hit, and
it will also be uncommon for two random noise hits to occur close enough
together to be paired into a space point. This means that
most random noise hits will be rejected when space points are
constructed as the first step of the pattern recognition algorithm.

   Although stereo strips give a position resolution perpendicular to
tracks' bend coordinate which is only moderate, the physically
important bend coordinate resolution -- which determines the momentum
measurement -- is essentially as good as if the two planes of strips
had been placed completely perpendicular to the bend coordinate.
This behaviour can be derived as follows.

   Consider strips with gaussian resolution $\sigma$ which are
oriented at (small) stereo half-angles $\alpha$ to the z-axis, in a
local coordinate system with the wafer in the
x-z plane. Using polar coordinates, $\rho=x^2 + z^2$ and
$\theta = \tan^{-1} (\frac{x}{z}$), one can write the following
formula for the negative log likelihood distribution, $-L$, of the
true position of a hit which is measured to be at the origin:

\begin{eqnarray*}
-L & = & (\frac{\rho}{\sigma})^2
          \{ \sin ^2 (\theta - \alpha) + 
                          \sin ^2 (\theta + \alpha) \}   \\
   & = &  2(\frac{\rho}{\sigma})^2
          \{\sin ^2 \theta \cos ^2 \alpha +
                          \cos ^2 \theta \sin ^2 \alpha \}  \\
   & \simeq & \frac{2(1-\alpha^2)}{\sigma ^2} x^2 +
                              \frac{2 \alpha ^2}{\sigma ^2} z^2, \\
\end{eqnarray*}
where terms of order $\alpha ^4$ and higher have been neglected in the
final step of the derivation. The likelihood function is seen to have
an elliptical distribution with the major
axis along the z direction, corresponding to Gaussian resolutions
of $\sigma (1-\frac{\alpha ^2}{2})/ \sqrt{2}$ and
$\sigma / (\sqrt{2} \alpha)$ for the x and z views, respectively.
Since $\alpha \ll 1$, the resolution in the x view is very close to
the optimal value, $\sigma / \sqrt{2}$, which would be obtained if
both layers of strips were oriented parallel to the z-axis.

   From the above discussion, the
choice of stereo angle can be seen to be a trade-off between simpler
pattern recognition and degraded resolution in the z coordinate.
In principle, the pattern recognition capabilities of the tracker
can be optimized by using different stereo angles in different
regions of the tracker. For example, smaller stereo angles could
be used in the forward regions of the mid-radius shells, where the
track density is greater than in the central region. In general,
a reasonable choice for stereo angle might be similar to the 2
degree angle which will be used for the tracker upgrade of the
D0 detector at the Fermilab Tevatron~\cite{dzero}.

\section{Silicon Microstrip and Pixel Detectors}

   This section presents general comments on the likely design
parameters for the silicon microstrip and pixel detectors, beginning
with a discussion on the microstrip detectors.

   The design of the silicon microstrip detectors can draw on
experience from many successful vertex detectors in today's
collider experiments, some of which have had design constraints
at least as demanding as will be imposed at the TLC. For the
microstrip detectors, the only new feature for the CT is the
logistics challenge of maintaining quality control over a larger
surface area of detectors than exists today.

   In spite of the large surface area of microstrips, it should be
noted that the approximately $2 \times 10^7$ channels of microstrips
is still considerably less than the of order $10^8$ channels of pixels
required for the vertex detector. The VXD3 vertex
detector~\cite{damerellpaper}, which has
just begun taking data in the SLD collider detector, already
incorporates $3 \times 10^8$ channels of CCD pixels.
(To get a feel for the numbers,
$2 \times 10^7$ microstrip channels corresponds to 100 ${\rm m^2}$
of 10 cm long microstrips with a 50 micron readout pitch, while
1 ${\rm m^2}$ of ${\rm 100 \mu m \times 100 \mu m}$ pixels is
$10^8$ channels.) Both the microstrip and pixel detectors will
incorporate front-end readout chips to sparsify (via zero
suppression), multiplex and buffer the channel signals, thus reducing
the output load to a more manageable level for the downstream
electronics~\cite{cmstp,atlastp}.

    For the best possible performance, it is likely that the
silicon microstrip detectors and much of the associated
electronics would be custom-designed for this tracker. This is
relatively normal for collider detectors -- taking advantage of some
of the techniques and facilities used for commercial silicon chip
design -- and is certainly justified by the large wafer area required
for this tracker.

    Many of the design parameters of microstrip detectors
have become relatively standardized. A wafer thickness of
300 microns is the norm, and a minimum strip read-out pitch of 
50 microns is fixed by the size of today's front end electronics.
A strip pitch of 25 microns is often used, with the ``floating
strips'' between the read-out strips used to collect charge
by capacitive coupling to the adjacent read-out strips. This
technique gives an improved interpolation of track positions
when combined with charge weighting of the signals on the
read-out strips. High momentum tracks will pass through
the wafers at near normal incidence and, for such tracks, point
resolutions of about 8 microns are typical for experiments
using this configuration. The length of the wafers will be
determined by a compromise between reducing the number of
electronics channels and the degradation of readout
signal-to-noise due to increasing strip capacitance. Strip
lengths from 6 to 12 cm have become fairly common, and the BaBar
silicon vertex tracker~\cite{babar} intends to use strips as long as
24 cm.

    Conventionally, single-sided silicon detectors have read
out the positively-charged ``holes'' produced by the ionizing
track, rather than the ionization electrons. In a 4 Tesla magnetic
field, the holes drift through the silicon at a 7 degree Lorentz
angle to the electric field direction. Although this is probably
not serious anyway, it can be corrected for
by tilting the detectors slightly in the azimuthal direction.

   The magnetic field poses much more serious problems for
double-sided read-out schemes, where the ionization electrons
are also read out from the back surface of the wafer. Electrons
experience a much larger Lorentz angle than holes -- 33 degrees
at 4 Tesla -- due to their greater mobility. The detectors
cannot be tilted to simultaneously satisfy the Lorentz angles
of both holes and electrons, and this level of transverse
drift might be expected to introduce resolution tails and
somewhat degrade the measurement of charged tracks at normal
incidence to the wafer. Double-sided detectors are also more
difficult and expensive to manufacture, currently costing
about three times as much as single-sided detectors.

    These disadvantages suggest that back-to-back single-sided
wafers are to be preferred for the CT's stereo layers,
despite the additional material this introduces: one extra
wafer thickness, which is 300 microns or 0.3 percent of a
radiation length.

    Given the size of the CT, it is likely that the front-end
electronics would be mounted on the wafers, as opposed to
carrying the raw signals out of the tracker on flexible fanout
circuits. Several types of existing front-end chips, typically
with 64 or 128 readout channels per chip, already offer a variety
of attractive read-out options for the CT~\cite{readout}. To reduce the rate of
data flow, front-end chips can store the amplified signals from
each strip in an analogue pipeline, to be read out only on the
receipt of a level-one event trigger. Both analogue and digital
read-out options are available, using either electrical or optical
links. Dedicated fast trigger channels can also be included,
perhaps with coarser granularity.

    Regarding the time constraints on the electronics, the
TESLA design's 707 ns beam-crossing interval places only relatively
moderate demands on the read-out electronics for the silicon strips,
but the 6 ns interval for S-band is approaching the smallest
interval at colliders planned for the near future~\cite{pdg}:
2 ns for the KEK B factory. For this
option, the read-out would probably have to integrate over a few
beam crossings.

    The main cause of radiation damage to the detector wafers will
be neutrons of energies around 1 MeV~\cite{schulte}, mostly originating
from the area of the final focus magnets. The vertexing layers will
be exposed to of order $3 \times 10^9$ neutrons per square centimetre per
year of operation~\cite{schulte} (within large uncertainties), and
the silicon strips to less than this. This is considerable, but,
for comparision, is still 4 orders of magnitude below the dose rate
expected at the LHC vertex detectors.

    In general, the read-out electronics can be made more robust
against radiation damage than the substrate itself~\cite{damerellpaper},
and it can even be designed to compensate for radiation-induced
changes in the signal from the silicon~\cite{cmstp,atlastp}.
Radiation-hard electronics
will be required for the CT, but won't have to cope with the level of bulk
damage to the silicon substrate that is expected for the detectors
at the LHC and that imposes stringent requirements
on the LHC electronics. There is also no strong motivation to use a
material other than silicon as the detector substrate.

   In short, the technology for silicon strip detectors is already
well established, and the detector design for the TLC will be geared
towards optimizing the resolution and reliability of the detectors.

   In contrast to the established pedigree for microstrips, there
are no pixel detectors in operation today that satisfy all of the
performance requirements of the CT vertex detector. However, rapid
progress is being made in this field and it is a reasonable
extrapolation to assume that such devices will become available
within the next couple of years.

   Pixel detectors come in two varieties: charge coupled devices
(CCD's) and active pixel sensors (APS's).
(See reference~\cite{damerellpaper}
for detailed descriptions of both pixel detectors and microstrip
detectors.)
 Each of the two types has advantages and disadvantages.
CCD's can be made thinner and with smaller pixels, giving them an
important advantage over APS's in position resolution. CCD's using
20 micron x 20 micron pixels can achieve outstanding point resolutions of
3.5 microns~\cite{damerelltalk}, while APS's might have trouble coming within
a factor of two of this figure.

   CCD's are also cheaper than APS's, and have the advantage of
several years of successful data-taking experience with the VXD2 and VXD3
vertex detectors in the SLD experiment. To balance this, APS's have already
been successfully used in a fixed target experiment at CERN and have
now begun data-taking in the forward region of the DELPHI vertex
detector. Also, a vigorous research and development program for APS's
is being carried out for the ATLAS~\cite{atlastp} and CMS~\cite{cmstp}
vertex detectors at the LHC, both of which will have similar areas to
the CT vertex detector and must meet much more stringent design demands.

   The Archille's heel of CCD's is their intrinsically slow
read-out. For both the TESLA and S-band accelerator options,
the CCD's would unavoidably collect background hits from
some few tens of beam crossings around the triggered event.
Also, some amount of dead-time (less than 30\%) might be introduced
for CCD's with the TESLA option. 
(See~\cite{damerelltalk} for CCD read-out strategies at the FLC.)
CCD's are also more vulnerable to radiation damage than either
APS's or microstrips. The expected neutron background at the
TLC is right at the limit of what is tolerable for today's
CCD detectors.

   In summary, many attractive design options already exist for the
silicon microstrip detectors, while the contrasting advantages of CCD's
and APS's and the rapid advances in both technologies suggest
that it would be sensible to delay the choice of pixel technology
for as long as is practical.


\section{Pattern Recognition}

   The task of doing track-finding using a few widely spaced,
precise coordinate measurements is familiar at test-beam lines
and fixed target experiments but relatively new for colliders.
This section describes some simple studies which give confidence
in the pattern recognition performance of the CT.

    The CT measures space points in 10 concentric layers plus
one additional sampling at the outermost layer, giving 11-plus-10
samplings in/out of the bend plane. Of other trackers at existing
and proposed collider experiments, CMS~\cite{cmstp}
has the most similar pattern recognition
strategy, with approximately 12-plus-6 or 12-plus-7
samplings (depending on the track direction). It should be
noted that the pattern recognition task is expected to be much
more demanding for CMS than for the CT, with a hit density
expected to be larger by up to two orders of magnitude and with
lower detector efficiencies due to radiation damage. In both
tracker designs the enormous number of channels -- of order $10^8$
pixel channels and $10^7$ microstrip channels -- would be expected
to be well suited for reconstructing high multiplicity events.

    Another sparse-sampling tracker design for a collider experiment
is the BaBar silicon vertex tracker~\cite{babar}, which is intended
to perform stand-alone tracking of low momentum tracks. The BaBar
design has only only 5-plus-5 samplings in/out of the bend plane,
reflecting the less dense track environment at the lower energy
PEP-II electron-positron collider.

    From the above discussion, the CT design appears to be more
conservative and robust for its pattern recognition task than either
the CMS or BaBar trackers. To check this assertion, a simple track-finding
software package was developed and tested on Monte Carlo-generated
events to give a more quantitative assessment
of the the pattern recognition performance of the CT~\cite{ucltalk}.

   The algorithm searches for track ``3-seeds'' consisting of triplets
of space points in any 3 distinct detector shells, where the space
points come either from pixel hits
or from matched pairs of hits in the double layers of stereo microstrips.
Unmatched hits in the silicon microstrips are simply ignored --
which is obviously not optimal. Pairs of 3-seeds with 2 space
points in common are then merged and tested to see if the 4
space points form a new, larger track seed (a ``4-seed''). Complete track
candidates are grown by continuing this process of seed-merging.
 
   The algorithm was tested on a previous design iteration of the
CT which incorporated only 8 shells, as opposed to the 10 shells
of the current version, and only 8-plus-8 samplings in/out of
the bend plane. Using this geometry, the pattern recognition
code was run on two ``toy'' events containing 100 isotropic muons
with momenta of 1 GeV/c and 2 GeV/c, respectively, and also on a
more realistic 2-jet event, at 500 GeV centre-of-mass energy, which
was generated by the PYTHIA event generator.

   The isotropic events were intended to test the performance of the
algorithm on high multiplicity events with soft tracks, for which the
large bend angles and large MCS deviations make the pattern recognition
more difficult than for stiff tracks. The 2-jet event gives a somewhat
complementary test, with tightly collimated tracks of higher momentum.

   For more detail on these test events, the tracks were
propagated through the tracker using the GEANT detector simulation software
package. Realistically simulated~\cite{schulte} noise tracks from
incoherent beam electron pairs were overlayed on the event. The
simulation included realistic Coulomb scattering (Moliere theory),
Gaussian resolution smearing of 5 microns r.m.s. and 2 percent
detector inefficiency per pixel or microstrip layer. Delta rays were
generated down to a 1 MeV threshold, but other interactions and decays
were not considered. In addition, ghost hits due to ambiguities
in the stereo strips were not included.

     Even for this non-optimal algorithm, the study found an
acceptable track-finding
efficiency of 99.6 percent: one track lost from the fiducial volume
out of 225 tracks with momenta of 1 GeV/c or greater. (The lost track
was from the event with 100 isotropic 2 GeV/c muons.)

    Further pattern recognition studies, using higher track statistics
and a more complete physics model, are needed before definitive
conclusions can be drawn. However, it is reasonable to assume that the
change to 10 detector layers along with improvements in the pattern
recognition algorithm should provide extremely efficient and robust
pattern recognition.

\section{Mechanical Issues}

     Most of this section addresses the mechanical support structure
of the CT, giving an overview of the support structure and then
describing a finite analysis computer study of the rigidity of the
structure. The section ends with a rough calculation of the space needed
to route electronic and cooling services out the ends of the tracker.

     Today's vertex detectors are stable at the level of a few
microns during data-taking runs. A similar level of stability
must be achieved over the much larger dimensions of the CT if
the momentum resolution of the tracker is not to be degraded.
Finite element stress calculations indicate that such a level
can be achieved using 3-dimensional support structures of carbon
fibre composite (CFC), as discussed below.

   The crucial property of CFC is its extremely small temperature
expansion coefficient: as low as $2.10^{-7}$ per degree centigrade --
which is roughly two orders of magnitude lower than metals. This
allows the construction of metre-scale CFC structures which have
an intrinsic thermal stability the same as, or better than, that
of the few-centrimetre-scale beryllium support structures currently
used in many vertex detectors.

   With a Young's modulus of up to 200 000 ${\rm N/mm}^2$ (the same
as steel) and a radiation length of about 24 cm, CFC also has
the stiffness required for building a light structure that won't
add more than about one percent of a radiation length to the tracker
mass. In addition, it is radiation hard and is easy to handle and machine.

    With the thermal behaviour already taken care of, any
other mechanical stress in the structure can be
minimized by mechanically isolating the tracker. To achieve
this, the CT will be supported at both ends on the tungsten
mask, with one end fixed to the mask but the other end mounted
on rollers to allow free movement parallel to the beamline.
Any movement of the tungsten masks will move the CT as a
whole, without distorting its shape or degrading its resolution.
(It should be noted, however, that transverse motion of the masks
can still degrade the knowledge of the beam-spot position.) We will
return to describe a study of the CT's mechanical rigidity,
after first giving details about the structural layout.

   The large-scale structure of the CT consists of six modules.
Firstly, the tracker splits in two, in a vertical plane through its
long axis, to allow assembly and dis-assembly around the
beam-pipe and tungsten radiation masks. Then, each of the halves is
radially divided into three
concentric sections: the innermost section consists of the 6-layer
vertex detector and the other 2 sections each support 2 of the
4 large shells of detectors  -- shells 7 and 8 (9 and 10) for the
smaller (larger) of the modules, counting outwards from the IP.

    The ends of the six CT
sections can be very rigidly joined to conical support
structures lying directly over the tungsten mask. This is
outside the active volume of the tracker, so thick CFC struts
can be used without degrading the tracker performance.

   The mechanical design of the vertex detector can be
patterned after that of the (similar) BaBar silicon vertex
tracker~\cite{babar}. The BaBar tracker will use six mechanical
modules, forming sextants in the azimuthal direction, which each
contain several silicon detectors glued to CFC
support beams. Some of the support beams will run within detector
layers and others run radially between layers, and the
modules will be attached to CFC support cones at the ends
of the vertex detector.

   The two large modules in each half of the CT consist of
3-dimensional structures of CFC struts, with triangular tetrahedra
used as the basic structural unit. This ensures that the structure
cannot flex in any direction without either compressing or stretching
some of the struts. Both types of modules are tiled with the silicon
detectors on both their inner and outer surfaces. In more detail,
longitudinal CFC ribs, which are glued to the space structure,
will support the semi-circular arcs of detectors, each of which
will consist of several CFC modules containing a convenient number
of detectors.

   A mechanical design study has been made for an example layout of
the larger of the two types of 3-D modules~\cite{ucltalk}, since
this is assumed to be the most demanding part of the CT mechanics.
(To be precise, the layout was for an earlier iteration of the
CT, which had the slightly smaller radius of 85 cm. However,
the results from the study should also be approximately correct
for the current design.) The example module comprised
approximately 120 CFC struts along the ellipsoidal surface of shell 9,
160 struts along shell 10 and approximately 160 further struts
joining the shells. The cross-sections of these 3 classes of struts
were 1, 4 and 1 square centimetres, respectively, which corresponds
to an average of less than 1 percent of a radiation length per shell.

   It is difficult to specify and quantify the exact stresses
on the tracker during data-taking. Since the CT is mechanically
isolated, however, it is reasonable to assume that the
time-dependent stresses should be much less than the constant
gravitational stress from the weight of the tracker. Hence,
the sag of the structure under its own weight acts as a very
conservative benchmark for how much it could distort during data
taking.

\begin{figure}[tb]
\begin{center}
\mbox{\epsfxsize=6.8cm\epsffile[0 60 565 500]{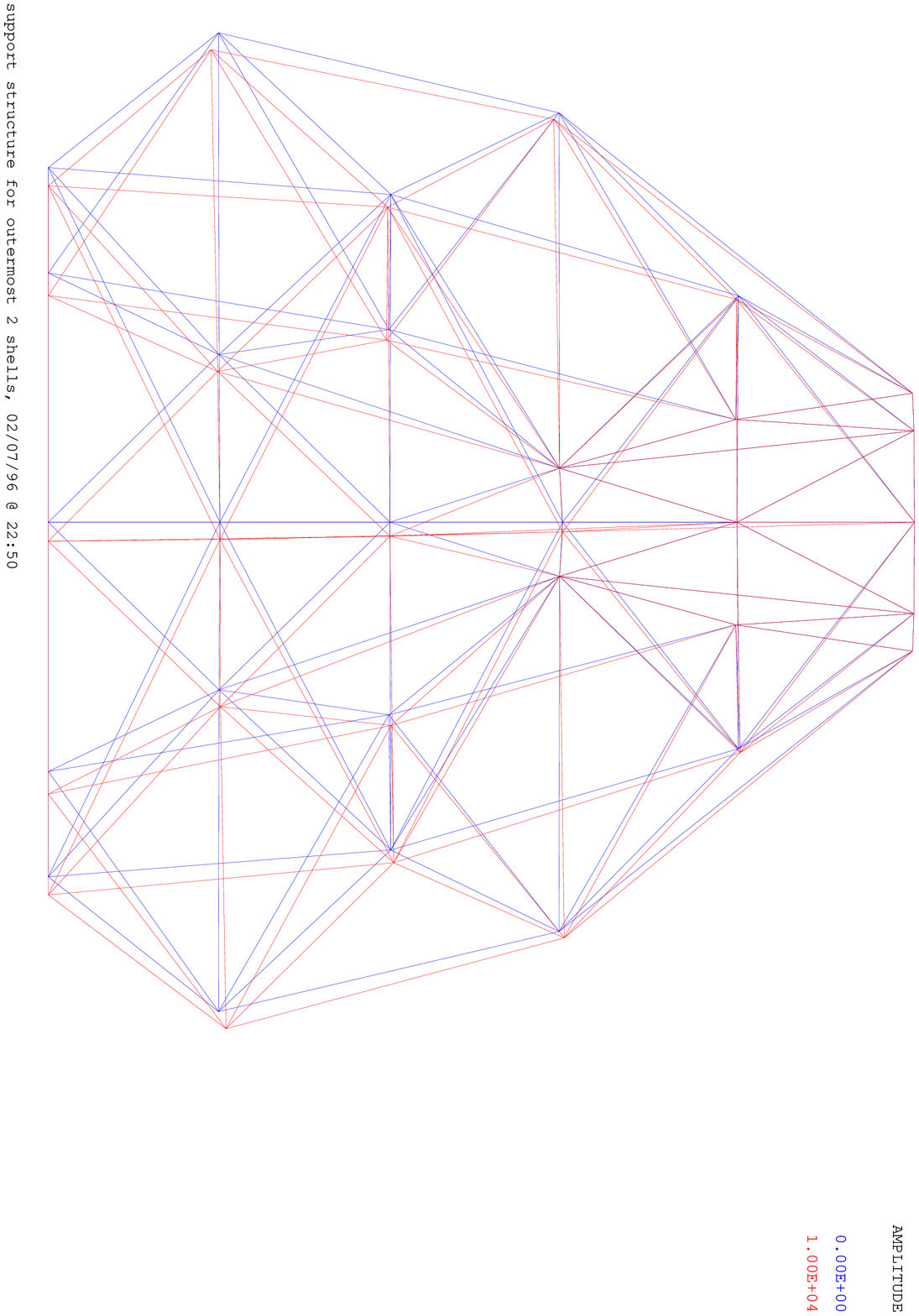}}
\end{center}
\caption{
{\small \bf
   An example design for the large 3-dimensional mechanical support
module that was used for finite analysis stress
studies. Half of the module is shown, viewed from a position
horizontally displaced from the IP in a direction perpendicular
to the beam-line. The IP is half-way up the left hand side of the
figure, with the unseen half of the module being the mirror image
of the displayed half in a vertical plane through the IP. The two
sets of slightly displaced lines show the structure a) without
gravitational loading and b) with 10 000 times the deformation
produced by gravitational loading.
}
}
\label{fig:castem1}
\end{figure}

\begin{figure}[tb]
\begin{center}
\mbox{\epsfxsize=6.8cm\epsffile[0 60 565 500]{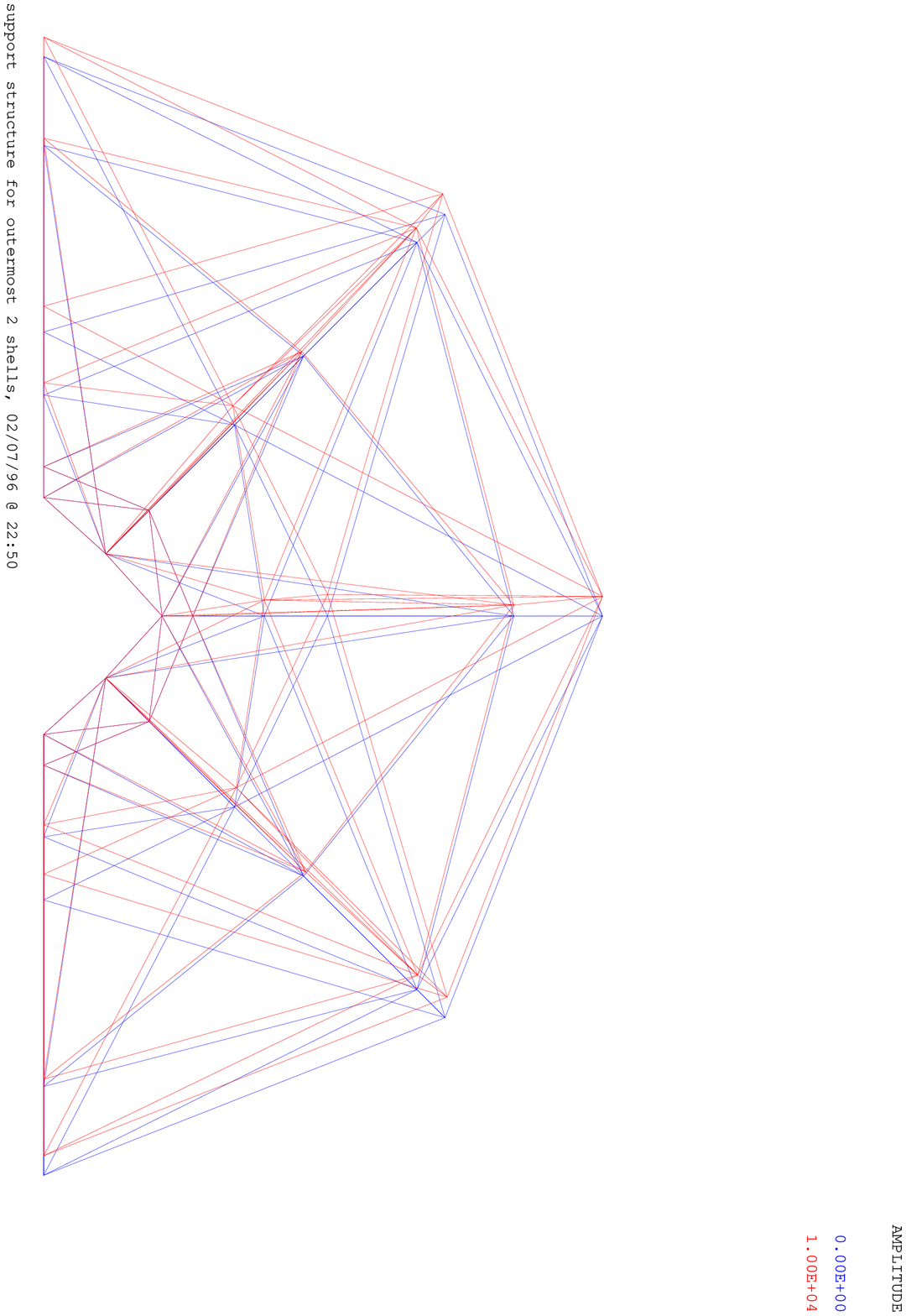}}
\end{center}
\caption{
{\small \bf
   Same as figure 3, but in a view looking along the beam-pipe,
such that the beam-pipe enters into the page half-way up the left
hand side of the figure.
}
}
\label{fig:castem2}
\end{figure}

   Finite analysis calculations for the example module,
using the CASTEM2000~\cite{castem} software package,
determined the maximum graviational sag
of the structure to be approximately 5 microns for an assumed
weight of 400 kg. (The assumed weight is probably too large by
at least a factor of two.) Figures~\ref{fig:castem1}
and~\ref{fig:castem2} show two views of the example module,
including a greatly magnified illustration of the nature of
the gravitational sag.

   The results of the finite element calculations suggest strongly
that the tracker performance will not be adversely affected
by distortions in its large-scale mechanical structure.
If anything, an even lighter version of the 3-D mechanical
structure would probably give sufficient rigidity.

   As a final mechanical design issue, sufficient space must be
provided for the routing of cables and cooling
pipes out of the CT ends. The following crude calculation suggests
that this will probably not be a serious problem.

   The $2 \times 10^7$
channels of microstrip detectors will typically draw about 2 mW per
channel~\cite{cmstp}, or 40 kW in total. If APS's are used in the
vertex detector then they might be expected to consume of order
0.5 ${\rm W.cm^{-2}}$~\cite{cmstp,atlastp},
adding about 10 kW to the total power consumption. (CCD pixels would
require much less power~\cite{damerellpaper}.) From the Ohmic
heating law (${\rm P=IV}$, in obvious notation), this
power consumption corresponds to a current of 5000 A for an assumed
average voltage drop of order 10 V. From Ohm's law, ${\rm V=IR}$,
a total cross-section of only 10 ${\rm cm}^2$ of aluminium conductor
(resistivity of
$2.7 \times 10^{-8} \Omega . {\rm m}$ at 300 K) would suffice to
supply this current,  with a manageable voltage drop of about
1 volt every 7 m.

    The 10 kW power consumed by the front-end electronics will appear
as heat. In a realistic scenario, this could be removed by a
60 ${\rm cm}^2$ cross-section of water (which has a specific heat of
4.2 ${\rm J. K^{-1}. cm^{-3}}$) flowing at 1 meter per second and with
an average temperature rise of approximately 2 degrees centigrade.

    To obtain the total space requirements
for these services, the aluminium and water cross-sections must
be multiplied by factors to account for cladding, load safety
factors, additional services and packing efficiency. A crudely
estimated total space requirement of around 500 ${\rm cm}^2$ 
would result from safe multiplicative factors of 20 to 30 for
the aluminium cross-section and 4 to 5 for the cooling water. This
area corresponds to a 2 cm--thick annular aperture,
with 20 cm radius, at each end of the tracker.
Thus it appears that the routing of services from the tracker
will make only fairly modest demands on the tracker geometry.

\section{Alignment}

    Knowledge of the wafer positions at the level of one or two microns
is required to keep pace with the excellent point precisions of the
silicon detectors. This demanding specification seems to be
achievable, given that the solid state technology and the stiff
mechanical structure ensure that the alignment constants won't change
with time.

    Precise mounting of the wafers, followed by an optical survey,
fixes the radial alignment constants but cannot give the required accuracy
in the transverse directions. The additional information must come
from high statistics fitting of charged tracks, which, in turn,
implies the need for dedicated calibration running on the ${\rm Z}^0$
resonance, at 91.2 GeV centre-of-mass energy.
(It should be noted that a strong physics case can also be made
for a high statistics ${\rm Z}^0$ run.) We now give a general overview
of how the alignment could proceed.

    The alignment procedure can be conceptually divided into two
parts, with one part concentrating on local coordinate distortions
and the other part concerning the more global alignment.
The local alignment of neighbouring wafers within shells is attained by
iteratively adjusting the transverse alignment constants to minimize
the sum of the track chi-squares. This should be greatly aided, in this
particular tracker, by the precisely known, time-independent geometry
of the silicon wafers and by their overlap in both views.

   Although effective for removing local misalignments, it is
well known that a chi-squared minimization cannot completely
determine the wafer positions, even in principle. The particular
remaining global distortions which can degrade the momentum resolution
of the tracker are rotations of the mid-radius shells relative to
the inner and outer shells. In this case, the momenta of one sign
of tracks will be under-measured and the opposite sign over-measured,
but without necessarily increasing the tracks' chi-squared.

   The second part of the track-based alignment procedure, which removes these
distortions,uses the sample of back-to-back pairs of 45.6 GeV/c muons
from decays, at rest, of the ${\rm Z}^0$. Consider two such muons from
different ${\rm Z}^0$ events, with opposite signs and which happen to have
the same coordinates in the innermost and outermost shells. It is
clear that these muons will then
have coordinates in the intermediate shells which are
symmetrically displaced on either side of a straight line.
Thus, a simple average of the hit positions gives an un-biased
alignment for the intermediate shells. This general procedure
assumes only that the ${\rm Z}^0$'s are produced, on average, at rest
and is true even in the presence of initial- and final-state
radiation and a forward-backward muon asymmetry.

   The order-of-magnitude integrated luminosity required for each
alignment run can be calculated under the reasonable assumption
that the final alignment step requires the most luminosity, since
it can only use the dimuon channel. The total number of muon
pairs required for the alignment can be estimated as the product
of two terms, one for the alignment in a given direction
and the other to take account of the effective number of different
directions in which the final alignment step must be done, as follows.

   Under the idealized assumption of monochromatic muon tracks of
very high energy, only one track of each sign would be required to
determine the alignment in any given direction to an accuracy
equal to the point resolution. (To be precise, the alignment
uncertainty would actually be a factor of $1/\sqrt{2}$ better than
the point resolution.) More realistically, multiple scattering is
still important for the 45.6 GeV muon tracks from ${\rm Z}^0$ decays
at rest, and additional statistics might also be
needed due to a non-zero spread in the transverse momenta
distributions. Since we also wish the alignment to be determined
much more accurately than the point resolution, we conservatively
assume that 100 muons of each sign would determine the alignment
well enough in any specified direction.

   It is to be expected that independent
misalignments would generally exist at widely spaced polar angles,
but, on the other hand,
but it is likely that any azimuthal variation would already
have been corrected for by requiring the momentum spectra
of all charged tracks to be azimuthally symmetric. A fairly
conservative assumption is that each of the bands of wafers
in the outer shell (37 bands, for 10 cm long wafers)
would effectively need a separate alignment constant.
This corresponds to of order 100 ``independent directions''.

   Multiplying the factors predicts that of order 10 000
muon pairs would suffice to do the final alignment step
to much better than the point precision of the tracker.
Since the cross-section-times-branching-ratio for dimuons
at the ${\rm Z}^0$ resonance is about one nanobarn, this corresponds
to an integrated luminosity of order 10 inverse picobarns,
or $10^4$ to $10^5$ seconds of running at realistic FLC
luminosities of $10^{33}$ or $10^{32}$
${\rm cm}^{-2}{\rm s}^{-1}$.

   In summary, it appears that the CT could be accurately
aligned with only a short run at the  ${\rm Z}^0$ peak, say,
once every year.

\section{Central Tracker Performance}

   This section uses heuristic arguments to derive simple expressions
for the vertexing and momentum resolutions of the CT as a function of
polar angle. A slightly more rigorous estimate of the momentum resolution
for high momentum tracks is also presented.

   Considering first the vertexing performance,
a simple parameter which gives a good indication of the performance
of the vertex detector is the width, $\sigma_{ip}$,
of the measured impact parameter distribution for tracks originating
from the IP. (More precisely, $\sigma_{ip}$ is the width in either
the r-phi or z projection.)

The impact parameter uncertainty will have contributions from both
the point resolutions of the pixel detectors and from multiple coulomb
scattering (MCS) in the beam-pipe
and detector material, so a standard general form is:
\begin{equation}
      \sigma_{ip} = a(\theta)  \oplus  b (\theta) / p ,
\end{equation}
where p is the track momentum, $\theta$ the polar angle of the track,
and the terms in $a$ (point resolution term) and $b$ (MCS term)
are added in quadrature. The parameters may be estimated by
considering the asymptotic limits of extremely high momentum and
low momentum tracks, where the first or second terms will dominate,
respectively.

   MCS can be neglected in the high momentum limit, so the direction
of tracks can be very precisely measured using the full lever-arm of
the outer tracker. Out of the bend plane, the outermost position
determinations will have uncertainties of order 100 microns (depending
on the stereo angle chosen for the microstrips) at a lever-arm of about 1
metre, corresponding to an angular uncertainty of only 0.1 milliradians.
In the bend plane, resolutions of a few microns for both the mid- and
outer-radius shells should give an even better angular measurement,
even allowing for the curvature of the track in the magnetic field.
For determining the impact parameter, these angular uncertainties
will be multiplied by the extrapolation distance from the vertex
detector to the IP -- a few centimetres -- which will give a
contribution of a few microns to $\sigma_{ip}$.

   A comparable contribution to $\sigma_{ip}$ in the high momentum
limit will come from the
point resolution of the 6 planes of vertex detector (i.e. $1/\sqrt{6}$
times the point resolution of an individual plane). Hence, on adding
the two contributions in quadrature, one might expect the high-momentum
limit of the impact parameter resolution to be of order:
\begin{equation}
a(\theta) \simeq 5\;\; {\rm microns},
\end{equation}
with little dependence on polar angle.

    To estimate the MCS term in $\sigma_{ip}$, consider that
very low momentum tracks will scatter through a relatively
large angle at the innermost layer of the tracker. The average
angle is given by~\cite{pdg}:
\begin{equation}
\theta_{MCS} \simeq 0.0136 \sqrt{X_0} / {\rm p(GeV/c)},  \label{eq:mcs}
\end{equation}
where $X_0$ is the fraction of a radiation length of material
contained in the beampipe plus innermost layer of pixels.
The average
extrapolation error to the IP is the product of $\theta_{MCS}$
and the distance to the first layer. At normal incidence,
$\theta =90^{\circ}$, this distance is the radius of the
first layer:
\begin{equation}
b = \theta_{MCS} \times {\rm (radius\;of\;layer\;1)}.
\end{equation}
Assuming a radius of 2 cm and a reasonably
conservative thickness of
1 percent of a radiation length~\cite{atlastp,damerellpaper} (this is for
APS's -- CCD's can be made considerably thinner) gives:
\begin{equation}
b(90^{\circ}) \simeq 30\; \mu{\rm m}.{\rm GeV/c}.
\end{equation}

   For smaller angles of incidence,  $b(\theta)$ will become larger
due to the increased extrapolation distance
to the IP and to traversing a greater thickness of material.
In the central region, where the innermost layer is a barrel,
it is easily seen that the angular dependence goes as~\cite{damerellpaper}:
\begin{equation}
b(\theta) \simeq 30/( \sin \theta )^{3/2} \; \mu{\rm m}.{\rm GeV/c}.
\end{equation}
   In moving to smaller polar angles, the resolution will undergo
discontinuities at the ends of shells 1 and 2 which are not reproduced
by this simple form. The equation is, however, a reasonable
starting point for simple studies of the physics potential of
the vertex detector.

    For physics analyses using today's vertex detectors, the
tails on the impact parameter distributions are often at least as
important as the widths of the distributions. Hopefully, the
resolution tails should be very much reduced for the CT vertex
detector, with 6 layers of pixels rather than 2 or 3 layers
of microstrips. (The SLD vertex detector has already demonstrated
the advantages of pixels for obtaining smaller, better-modelled
resolution tails~\cite{damerellpaper}.)

   On comparing the numerical values of the constants $a$ and $b$,
it can be seen that the MCS and point-resolution contributions
to $\sigma_{ip}$ are of similar size for track momenta of several
GeV/c. Since some of the tracks with lifetime information will
be in this momentum range, it is clear that both terms will be
relevant for physics analyses.
It also appears that the impact parameter resolution is good
enough to give a very high efficiency and purity for finding
beauty and charm vertices and even for flagging the tracks from
1-prong decays of tau leptons (86\% of all tau decays).
This can be shown using a heuristic argument, as follows.

   From relativistic kinematics, a relativistic particle
with lifetime-at-rest $\tau$ will survive for an average
distance $L \simeq \gamma c \tau$, with $c$ the speed of
light and $\gamma$ the relativistic boost. Also, the forward
hemisphere of the particle's rest frame is compressed,
in the lab frame, into a cone of half-angle $1/\gamma$.
Thus, the ``typical track'' in a ``typical decay'' will have
an impact parameter of $L/\gamma = c\tau$, independent
of the relativistic boost. The ``typical'' impact parameters
for B hadrons ($c\tau \simeq 450\; \mu {\rm m}$),
charm hadrons ($c\tau = 320\; \mu {\rm m}$ for ${\rm D}^+$, 130
$\mu {\rm m}$ for ${\rm D}^0$), and tau leptons
($c\tau = 91\; \mu {\rm m}$), are all seen to be much larger than the
several micron uncertainty in the resolution. This suggests that
identifying the ``typical'' decay topologies will not be too
difficult, even for tau decays. Instead, the challenge will be to
push for very high efficiency and purity by identifying those decays
with shorter-than-average lifetimes and small impact parameters.

    We now move on to a discussion of the performance of the CT for
measuring track momenta, again using simple heuristic arguments to
derive numerical expressions for the momentum resolution.
The general form of the momentum resolution equation is expected
to be~\cite{pdg}:
\begin{equation}
         \sigma (p_t) / p_t = c (\theta) \oplus d (\theta) \times p_t,
                                       \label{eq:genmom}
\end{equation}
where $p_t$ is the component of the momentum transverse to the B
field, and the constants $c$ and $d$ represent the contributions
from MCS and point resolutions, respectively.

    The easiest way to derive approximate values for the two
constants is to use the simplified model of three track position
measurements: one at each of the inner-, mid- and outer-radii.
(To be precise, the radius is defined in the plane transverse to the
magnetic field, so it is a function of the polar angle.)  
If the three measurements are labelled $f$, $g$ and $h$, respectively,
then the measured quantity which is related to the track momentum is the
sagitta, $S$, defined by:
\begin{equation}
S = (f+h)/2 - g.      \label{eq:sagdef}
\end{equation}

   Starting from the Lorentz force equation,
it is easy to derive the following expression for the sagitta in terms
of the magnitudes of the magnetic field strength ($B$), transverse
momentum ($p_t$), charge ($q$) and half-radius in the transverse
plane ($L(\theta)$):
\begin{equation}
S = L(\theta)^2 q B / 2p_t.   \label{eq:sag}
\end{equation}
(A small angle approximation has been used, and the form of the
magnetic force equation is that appropriate for MKS units: $F=qvB$.)
For a singly charged track, with $S$ and $L$ in metres, $B$ in Tesla, and
converting momentum units to GeV/c, this becomes:
\begin{equation}
S = 0.150 L(\theta)^2 B / p_t(GeV/c).
\end{equation}

    It follows that the value of $d$ can be evaluated from
\begin{equation}
d(\theta) = \sigma_{S,point} / (0.150 L(\theta)^2 B),
\end{equation}
with the uncertainty in the sagitta for high momentum tracks,
$\sigma_{S,point}$, given by the sum-in-quadrature of the
uncertainties in the terms of equation~\ref{eq:sagdef}:
\begin{equation}
\sigma_S = \sigma_f/2 \oplus \sigma_h / 2 \oplus \sigma_g.
\end{equation}

   The CT at normal incidence to the beamline has
$L \simeq 0.5$ metres, $B=4$ Tesla and $\sigma_{S,point}
\simeq 8$ microns, so
\begin{equation}
d(\theta=90^{\circ}) \simeq 5 \times 10^{-5}\;\; (GeV/c)^{-1}\;\;\;\;\;\;
                                                ({\rm estimated}).
\end{equation}

   As confirmation of this equation, a more precise
value,
\begin{equation}
d(\theta=90^{\circ}) \simeq  3.6 \times 10^{-5}\;\; ({\rm GeV/c})^{-1}
                                        \;\;\;\;\;\;({\rm fitted}),
\end{equation}
was obtained~\cite{bjkmom} for
d($\theta =90^{\circ}$), using  the MINUIT fitting software~\cite{minuit}
with the assumption of one hit per detector
layer and effective point resolutions (i.e. including misalignments)
of 8 microns per detector. (No beam constraint was used.)

    For the angular dependence of $d$, one can take advantage
of the simple dependence of the tracker geometry on polar angle:
the radial distance to each of the detector shells scales in
approximate proportion to the radial distance to the outermost
shell, $R(\theta)$, and the outermost shell is an ellipsoid:
\begin{equation}
            (Z(\theta)/A)^2 + (R(\theta)/B)^2 = 1,
\end{equation}
with $A=1.74$ m and $B=1.00$ m.
All of the above equations still apply, with
\begin{equation}
L(\theta) = R(\theta) / 2,    \label{eq:ltheta}
\end{equation}
so,
\begin{equation}
       d(\theta) = d(90^\circ)\{ \frac{R(90^\circ)}{R(\theta)} \} ^2.
\end{equation}
Using the explicit angular dependence of R for the ellipsoid,
\begin{equation}
1 / (R(\theta))^2 = \cot ^2 \theta / A^2 + 1 / B^2,
                                   \label{eq:rvstheta}
\end{equation}
and putting in the numbers for the CT gives the explicit
angular dependence:
\begin{equation}
       d(\theta) = 3.6 \times 10^{-5} \{ 0.33 \cot ^2 \theta  +1 \}
                                                  (GeV/c)^{-1}.
\end{equation}
This estimate is found to agree well with the fitted values for several
polar angles.

    The MCS term, $c(\theta)$, depends on the distribution of mass in
the tracker, so it is slightly more difficult to estimate. Material
near mid-radius clearly contributes much more to the momentum uncertainty
than material near the inner and outer radii. In the simplified
model, the uncertainty in the outer endpoint due to a scattering
centre between mid-radius and outer-radius is clearly just the average
scattering angle times the lever-arm distance to the outer radius:
\begin{equation}
\delta h = ({\rm lever\; arm})\;\; \times\;\; \theta_{MCS}.
                                         \label{eq:delmcs}
\end{equation}
(A corresponding equation holds for scattering centres inside
mid-radius.) For the worst case, a scattering centre of $X_0$
radiation lengths at mid-radius (lever-arm distance = $L$) one
gets, on combining equations \ref{eq:mcs}, \ref{eq:genmom} and
\ref{eq:delmcs}:
\begin{equation}
c(\theta=90^\circ) = 0.045 \sqrt{X_0} / (L(\theta).B).
\end{equation}
On putting in the numbers for the CT and using the reasonable guess
that the amount of scattering material around mid-radius corresponds
to 4 percent of a radiation length at mid-radius, one finds:
\begin{equation}
c(\theta=90^\circ) = 0.0045.
\end{equation}

    The simple geometry of the CT again allows an explicit
parameterization of $a(\theta)$. If $D(\theta)$ is the distance
from the IP to the edge of the tracker for polar angle $\theta$
then
\begin{eqnarray*}
c(\theta) & \propto &  D(\theta).D(\theta).\frac{R(\theta)}{D(\theta)}.
                   \frac{1}{R(\theta)^2} \\
          & \propto &  D(\theta)/R(\theta) \\
          & = &        1/\sin(\theta).
\end{eqnarray*}
The interpretation of the terms is as follows. The first factor of
${D(\theta)}$ comes from the increasing material at smaller
polar angle, using equation~\ref{eq:mcs} and the convenient and rather
conservative assumption that the average density of material is
proportional to $D(\theta)$. The second factor of $D(\theta)$ is the
lever-arm for scattering (equation~\ref{eq:delmcs}). The third term,
$R(\theta)/D(\theta)$, accounts for the fact that the scattering
angle depends on the track's total momentum, rather than just
the transverse component. The final term, $1/R(\theta)^2$,
is simply the decrease in the sagitta given by equations
\ref{eq:sag} and \ref{eq:ltheta}.
Hence,
\begin{equation}
       c(\theta)  \simeq 0.0045 / \sin(\theta).
\end{equation}

    To summarize this section, the impact parameter resolution
and momentum resolution are valuable indicators of the performance
of the CT. Using heuristic arguments, explicit values
for these parameters were estimated as a function of polar
angle:
\begin{equation}
      \sigma_{ip} \simeq 5 \oplus  30/( \sin \theta )^{3/2}
                           / p({\rm GeV/c})
                          \;\;\; {\rm microns} ,
\end{equation}
and,
\begin{equation}
         \sigma (p_t) / p_t \simeq 0.0045 / \sin(\theta)\; \oplus \;
        3.6 \times 10^{-5} \{ 0.33 \cot ^2 \theta  +1 \}
        \times p_t ({\rm GeV/c}).
\end{equation}

\section{Central Tracker Cost}

   The cost of the CT should be completely dominated by the cost
of the silicon detectors and associated electronics, with the
mechanical structure of the CT probably contributing less than
1 MDM (million Deutchmarks).

   For illustration, carbon fibre struts
suitable for the 3-D support structures cost about 250 Swiss Francs
per metre, with little dependence on cross-section~\cite{gerwig}.
At this price, the large 3-D structures supporting the two outermost
shells contain a total of only about 30 000 DM worth of carbon
fibre composite.

    Large uncertainties in the cost of solid state detectors and
associated electronics preclude a detailed cost estimate for the
CT. Fortunately, prices for these components are rapidly dropping
as the technology advances, so an estimate based on today's prices
is almost certain to be inflated. A conservative over-estimate is
obtained from the cost calculations used for the CMS~\cite{cmscost}
and ATLAS~\cite{atlascost} experiments. In particular, the radiation
environment at the LHC requires very high quality silicon wafers
and very specialized electronics, which will not be needed at
the TLC.

    Crudely scaling the total cost of the ATLAS silicon
central tracking, 31 MCHF (million Swiss Francs), by the ratio
of the detector areas -- 96 ${\rm m}^2$ for the CT and 42 ${\rm m}^2$
for ATLAS -- gives a microstrip cost of 71 MCHF. This cost includes
the detectors, 20 percent spare detectors, electronics for the
read-out, mechanical support and installation. Similarly,
scaling the cost of the ATLAS APS's, 13 MCHF, by the ratio of areas
(1.75 ${\rm m}^2$ for the CT and 4.1 ${\rm m}^2$ for ATLAS)
adds an additional 6 MCHF to the total cost. (Using CCD's instead
of APS's would somewhat reduce the vertex detector cost.)

Summing the contributions from the mechanical support structure,
silicon microstrips and pixels gives a total tracker cost of a little
less than 80 MCHF. As already discussed, this can be considered to be a
conservative upper bound and the true cost is likely to be
less than this.

\section{Detector Design for 500 GeV e+e- Collider}

   This rather long section presents an example of a
detector design which incorporates the CT described above
and which would be suitable for the proposed DESY
electron-positron collider or a similar accelerator. Besides
its intrinsic interest, this exercise demonstrates that the
CT design is compatible with an affordable, high performance
detector and also motivates some of the parameters used in
the CT studies -- such as the 4 Tesla magnetic field.
However, it should be emphasized that the detector design has
not been studied at the same level of detail as the central
tracker. It is quite conceivable that better options exist
for some of the other subdetectors.

\begin{figure}[tb]
\begin{center}
\mbox{\epsfxsize=11cm\epsffile[15 0 550 500]{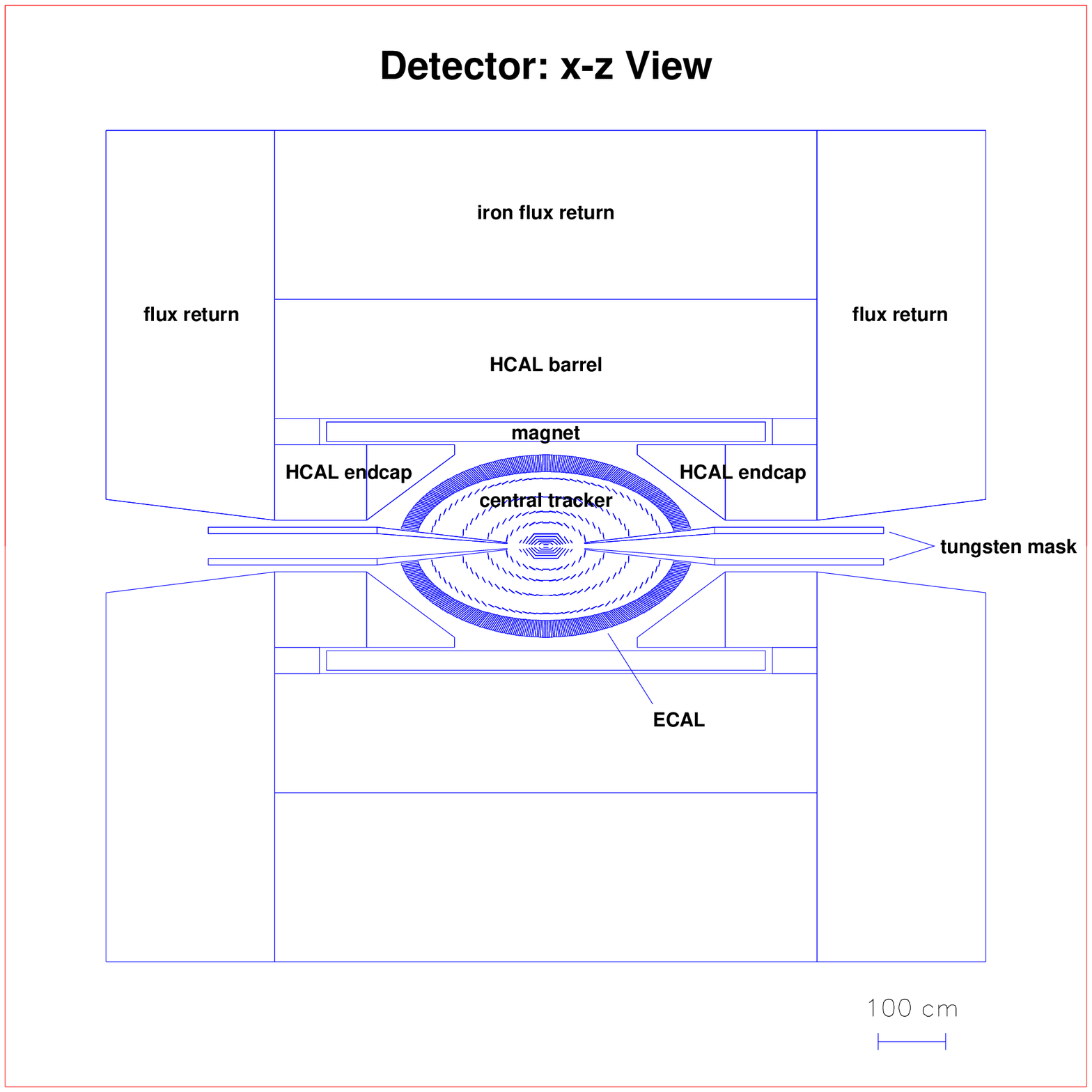}}
\end{center}
\caption{
{\small \bf
An example detector design which incorporates the all-silicon
central tracker. The view is a cross-sectional slice along the
beam-line.}
}
\label{fig:detr}
\end{figure}

   Figure~\ref{fig:detr} is an illustration of the detector.
The design is similar in some ways to the
CMS detector, which also uses a crystal electromagnetic
calorimeter (ECAL) and a 4 Tesla magnet. However, the central region
of CMS is less projective and CMS has the hadron barrel
calorimeter (HCAL) inside the magnet.

  Subsections 9.1 through 9.4 describe the subdetectors outside
the central tracker:
the electromagnetic calorimeter, magnet, scintillating
tile arrays and  hadron calorimeter. Subsection 9.5 gives
an overview of the detector performance, subsection 9.6
addresses the mechanical structure of the detector and
subsection 9.7 gives a rough estimate of the detector
cost, before the section is summarized in subsection 9.8.

\subsection{Electromagnetic Calorimeter}

    The electromagnetic calorimeter (ECAL) consists of an ellipsoidal
array of lead tungstate scintillating crystals, read out by
avalanche photodiodes. This choice is motivated by the excellent
energy resolution for electromagnetic showers:
\begin{equation}
\sigma {\rm (E)/E} = 0.5\% \oplus 2\%/\sqrt{{\rm E(GeV})} 
\end{equation}
and compactness (a radiation length of 0.89 cm and
Moliere radius of 2.2 cm) of these crystals.

   Additional attractive properties of lead tungstate are: fast
scintillation emission, good radiation hardness and a substantial existing
production capability. It should be noted, however, that achieving
a stable, uniform calibration for the crystal array is known to be
a difficult challenge. In this respect, the ECAL may benefit
from experience gained at the ZEUS detector, which is considering
installing a small array of these crystals as a forward electromagnetic
calorimeter. The CMS detector also intends to use lead tungstate
crystals for its ECAL, and is already undertaking a great deal
of research into the performance of these crystals.

    Reasonable dimensions for the crystals~\cite{brown} are
a depth of 28 radiation lengths (25 cm) and transverse dimensions
equal to the Moliere radius ($22 \times 22\; {\rm mm}^2$). The
crystals are oriented to be projective to the IP, and cover
an ellipsoidal surface just outside the central tracker, with
an inner surface which can be parameterized as:
\begin{equation}
(x^2 + y^2)/1.10^2 + z^2/1.94^2 = 1    \label{eq:ecalcoord}
\end{equation}
(with position coordinates $x,y$ and $z$ in units of metres and the
$z$ axis along the beam-line).

The position resolution for electromagnetic showeres is predicted to
be approximately ${\rm 2.2\; mm/\sqrt{E}}$~\cite{brown}.
Using equation~\ref{eq:ecalcoord},
it can be seen that this corresponds to an angular resolution of
$2.0\; {\rm mr}.({\rm GeV})^{1/2}$ in a direction perpendicular to the
beamline, improving smoothly to
$1.2\; {\rm mr}.({\rm GeV})^{1/2}$
at the calorimeter edge -- 120 mr from the beam direction.

   The mechanical support
of the crystals could proceed along the lines of that used for the L3
crystal calorimeter and the structure proposed for the CMS ECAL~\cite{cmstp},
with the crystals sandwiched between inner and outer support
frames. In the central region of this ECAL, each crystal is held in place
by an axial force applied to its back face using a
spring-loaded pusher. Each pusher encloses an avalanche photodiode,
thermometer and optical fibre for monitoring the crystal
calibration. A hemi-spherical plastic end-cap on the
front face of each crystal, which presses against the contoured
surface of the inner suppport frame, distributes the pressure
evenly over the crystal face. The crystals in the forward region
could be stacked as in conventional end-cap calorimeters.

   Following CMS, the inner support frame could consist of
carbon-fibre composite and hexel honey-comb, and the outer
support frame of aluminium. For this detector, the support frames would
be half-ellipsoids, with the ECAL split in two halves along
a plane perpendicular to the beam-line and passing through the IP.
The smaller ends of ellipsoids would
be attached to the tungsten mask and the larger ends of the
outer ellipsoids would be supported on rails attached to
the inner surface of the magnet. As in CMS, the crystals
themselves would contribute to the rigidity of the structure,
preventing the light inner suport frame from flattening under
the weight of the ECAL.

   As a specific example of a procedure for assembling the
calorimeter, each of the two inner support
frames could be constructed as a single piece but the outer
frames must be segmented to allow assembly and dis-assembly
of the crystal array and access to the electronics. The crystals
could be positioned and aligned beginning at the small ends of
the ellipsoids and proceeding row-by-row towards the middle of
the ECAL.

\subsection{Magnet and Scintillating Tile Arrays}

   Situated just outside the crystal calorimeter, the 4 Tesla
superconducting solenoidal magnet has an inner radius of 1.55 m,
half-length of 3.25 m and is 2.5 radiation lengths thick
(using the approximate formula given in~\cite{kircher}).
 The strong magnetic field improves the momentum
resolution of the tracker and also reduces the background in
the vertex detector by confining most of the soft beamstrahlung
tracks to remain inside the beam-pipe.

   The length of the magnet was determined from the magnet radius and
CT length according to the formula:
\begin{equation}
{\rm L_{magnet} = L_{tracker} + 2 \times R_{magnet} }.
\end{equation}
The motivation is that the magnetic field
inside a solenoid is relatively uniform up to about ${\rm R_{magnet}}$
from the ends. (This formula is approximately obeyed by the
CMS magnet, which will also be 4 Tesla.) This argument might
be somewhat conservative, and the magnet
can be shortened if calculations show that this won't
adversely affect the field in the forward region of
the central tracker.

   Because the magnetic field inside the solenoid is much stronger
than the saturation field for iron (1.5 to 1.8 Tesla) it should be
little affected by the iron flux return outside the magnet.
Thus, one can approximately calculate the fields along the z axis by
assuming a solenoidal field in a vacuum. A simple analytic calculation
predicts that at the maximum z of the inner tracker, 1.70 m, the
field along the beam
axis has fallen by only 8\%, giving good field uniformity for tracking.
Further out, fields are 3.5 T at 2m (the position of the innermost
quadrupole magnet for the S-band accelerator option) and 2.5 T at 3m
(the position of TESLA's  innermost quadrupole). This seems to be safely
below the 3 Tesla limit imposed for the TESLA option. In the S-band option,
the innermost accelerator magnet is assumed to be shielded by a
compensating solenoid, so the situation is less clear.

   Scintillating tile arrays, placed immediately inside
and outside the magnet, provide longitudinal and transverse sampling
of hadronic showers and should minimize the effect of the 0.6
interaction lengths of dead material.

   The tile dimensions could be as small as 10 cm by 10 cm, which
would correspond to about 5000 tiles in each of the two layers.
A typical configuration would consist of a wavelength-shifting fibre
embedded in the tile surface and connected to several metres of
clear fibre. Once outside the detector, the clear fibres could be
read out, for example, by either conventional multi-anode phototubes
or by Visible Light
Photon Counters (VLPC's) similar to those being installed for the
scintillating fibre tracker of the D0 experiment. VLPC's are a variant of
the solid state photomultiplier tube which operates in a liquid
helium cryostat. They provide greater than 70\% quantum efficiency
for visible light, with a gain of roughly 20 000 and a rate
capability of at least 10 MHz~\cite{dzero}.

    Assessing a cost of 300 DM per channel for the tiles and readout
gives a total estimated cost for the two tile arrays of around 3 MDM.

\subsection{Hadronic Calorimeter}  

    The hadron calorimeter (HCAL) consists of a barrel region outside
the magnet and
two end-cap calorimeters with slanted, partially projective front
faces to fit between the ellipsoidal ECAL and the magnet barrel.
The end-caps reach in to about 48 degrees from the beam direction.
Iron/scintillator sampling calorimetry is a likely choice for the
barrel HCAL (as in ATLAS), with copper/scintillator calorimetry in
the end-caps
(like the CMS HCAL, which is also inside a 4 Tesla field).

    Metal/scintillator sampling calorimeters are compact,
well understood and offer very good energy resolution. Because of
their relatively high densities, iron and copper are good choices
for the metal absorber. Further study is needed to optimize the
calorimeter depth and segmentation, and also to decide whether
a worthwhile improvement in performance would result from changing
to an even denser absorber, such as depleted uranium.

    Hadronic energy resolutions of $45\%/\sqrt{E}$
(ATLAS)~\cite{atlastp} and $65\%/\sqrt{E}$ (CMS)~\cite{cmstp}
have been estimated for the calorimeters themselves. Further
study is needed to estimate the performance when placed behind
both lead tungstate crystals and 0.6 interaction
lengths of magnet sandwiched between scintillating tile arrays.
(The lead tungstate crystals are a little more than one interaction
length, fully active but with an e/pi ratio of about 1.6.)

%
    Some evidence that the dead material in the magnet should not
seriously compromise the calorimeter performance is provided by the
large coarse-sampling calorimeters used in accelerator-based neutrino
experiments. The CCFR calorimeter uses liquid scintillation sampling
every 10 cm of iron -- that is, every 0.6 interaction lengths and 6
radiation lengths -- so each sampling length in the CCFR calorimeter
includes more material than in the entire thickness of the TLD magnet. In
spite of the coarse
sampling, the CCFR energy resolution measured in hadron test
beams~\cite{ccfr} is a moderate $85\% / \sqrt{{\rm E(GeV)}}$, with
only very small resolution tails for shower energies of order
100 GeV or above.

\subsection{Flux Return and Muon Chambers}

    The material in the flux return -- approximately 2 m of iron -- serves
as a hadronic filter for muon identification. The inner part can also
be instrumented as a hadronic tail-catcher if studies suggest this is
necessary.

   In general, substantially lower demands are placed on the
muon identification systems at electron-positron colliders
than at hadron colliders, due to the much reduced backgrounds
from punch-through hadrons.
The muons could be tracked using hermetic layers of drift tubes at
the outer radius and also, perhaps, at the mid-radius of the flux return.
The CMS chambers seem to provide a reasonable starting design:
8 (4) layers of drift tubes parallel (perpendicular) to the beam
direction plus resistive plate chambers for triggering and timing.

   The iron surrounding the HCAL provides a return path for the field
lines from the solenoidal magnet and also ranges out hadrons that
have penetrated the HCAL and would otherwise constitute a
background for muon identification. The minimum thickness of the
flux return is 12 interaction lengths, at z=0, which ensures that
the entire detector presents at least 20 interaction lengths to
hadrons originating from the collision region. In the forward
region, 2 m of iron is about what is necessary to contain the
field emanating from the solenoid. The iron cross-section at the
radius of the magnet is about 2.5 times the inner cross-section
of the magnet, corresponding to an average magnetic field, about 1.6
T, which is close to saturating the iron.

A detailed knowledge of the magnetic field in the iron would require
a finite element calculation on a computer.
However, it is expected that the detailed geometry of the flux return
should not seriously affect the magnetic field distribution in the
central tracker volume, since the magnet extends well beyond the
tracker and the 4 T magnetic field is well above saturation for iron.
Hence, the design of the flux return can be rather freely optimized
for muon detection.


\subsection{Detector Performance Overview}

   Operating at the energy frontier for electron-positron collisions
demands a detector with very good all-around performance, since it
is not known which new physics processes might present themselves.
Table~\ref{tab:detector performance}
summarizes the approximate estimated values, taken from
preceding sections, of
some of the most important resolution parameters associated with
each subdetector or event reconstruction task. The corresponding
values are also given for the ALEPH detector~\cite{aleph} at the LEP
accelerator, allowing a direct comparision between the TLD and a
detector operating at the current electron-positron energy frontier.
In spite of the very approximate nature of the comparision, it
can be seen that
the proposed detector performs significantly better in most areas
and is not worse for any of the parameters. This improvement can
mostly be attributed to recent progress in detector technologies.

\begin{table}
\centering
\begin{tabular}{|l|c|c|c|}
\hline
\hspace{1.0 cm}parameter  &  $\sigma_{\rm TLD}$    &  $\sigma_{\rm ALEPH}$   &
                       $\sigma_{\rm TLD} / \sigma_{\rm ALEPH}$ \\
\hline
CT Impact Parameter: &        &          &     \\
\hspace{0.5 cm} high mom. tracks  &  5 $\mu {\rm m}$  &  25 $\mu {\rm m}$ 
                                                         &  0.20 \\
\hspace{0.5 cm} low mom. tracks  &  30 $\mu {\rm m.GeV/c}$
                            &  95 $\mu {\rm m.GeV/c}$   &  0.32 \\
&&& \\
CT Momentum Resol.: &        &          &     \\
\hspace{0.5 cm} high mom. tracks & $0.36 \times 10^{-4}\; ({\rm GeV/c})^{-1}$
                                      & $6 \times 10^{-4}\; ({\rm GeV/c})^{-1}$
                                                           &  0.06 \\
\hspace{0.5 cm} low mom. tracks  &  0.0045 &  0.005   &  0.90 \\
&&& \\
EM Energy Resol.: &        &          &     \\
\hspace{0.5 cm} high energy limit          &  0.005  & 0.019  &  0.26 \\
\hspace{0.5 cm} stat. term ($1/\sqrt{E}$)  &  0.02 $({\rm GeV})^{1/2}$
                                            &  0.18 $({\rm GeV})^{1/2}$
                                            &  0.11 \\
EM Angular Resol.: &        &          &     \\
\hspace{0.5 cm} stat. term ($1/\sqrt{E}$)  &   2.2 ${\rm mr.({GeV})^{1/2}}$
                                            &   2.5 ${\rm mr.({GeV})^{1/2}}$
                                            &  0.88 \\
&&& \\
Hadronic Energy Resol.: &        &          &     \\
\hspace{0.5 cm} stat. term ($1/\sqrt{E}$)  &  0.65 $({\rm GeV})^{1/2}$ (?)
                                            &  0.85 $({\rm GeV})^{1/2}$
                                            &  0.76 (?) \\
\hline
\end{tabular} 
\caption{
{\small \bf
Approximate values for the uncertainties in some of the important
resolution parameters of the TLD. For comparision, the corresponding
uncertainties are also given for the ALEPH detector. The final column is
the ratio of the uncertainties for the FLD and ALEPH.
A value of 1.0 corresponds to equal performance, while
a smaller value represents an improvement of the FLC over ALEPH.}}
\label{tab:detector performance}
\end{table}

\subsection{Mechanical Structure and Detector Assembly}

   The mechanical design of the detector support structure is
intimately related to the procedure chosen for assembling and
dis-assembling the detector. Detailed planning of the assembly
procedure would involve a more complete knowledge of the
detector-machine interface than is now available. For the moment,
only a general outline will be given for a procedure which seems
to provide a good starting point for more detailed planning.

   The outer part of the detector -- the barrel region of the
HCAL, the flux return iron and the muon chambers -- splits vertically
in two along the major axis of the detector. These ``clam-shells''
are self-supporting and rest on rails which allow them to be
retracted in a horizontal direction perpendicular to the beam-line.
A sector at the bottom of the outer detector would remain in
place, supporting the magnet and inner detector. Special
arrangements would have to be made to access the electronics
channels in this sector -- this should not be a major design
problem.

   The inner detector is supported on the fixed rigid structure
provided by the magnet, as in the CMS design concept. Rails on
the inside surface of the
magnet allow the inner detector to be withdrawn in two halves,
with one half on each side of the IP. These rails can extend
beyond the ends of the magnet, parallel to the beam axis and
supported by the fixed bottom sector of the outer detector.

   The order in which the inner-detector
halves are withdrawn is important because the CT
is only fixed to one side of the tungsten mask. The opposite
side of the inner detector would be withdrawn first, with
the CT resting in its original position. As the un-anchored
side of the CT dropped off the receding mask it would be supported
by cables connected to the fixed half of the ECAL, whose purpose
is specifically to perform this task. Once
the first detector half has been withdrawn sufficiently, the
CT can be disassembled from the beam-pipe and removed for
maintainance.

  Each of the two inner-detector halves would consist of half of
the ECAL, one side of the tungsten radiation mask,
one of the HCAL end-caps and, optionally, end-cap sections
of the iron flux return and muon chambers. These would
be withdrawn as one piece, but a further set of rails
between the ECAL and HCAL would allow these sub-detectors
to be separated so that the ECAL electronics could be
accessed.

   To summarize, the
dis-assembly of the detector could proceed as follows:
\begin{enumerate}
\item Retract the clam-shells of the outer detector.
\item Disassemble the nearby beam-line.
\item Retract one half of the inner detector.
\item Dis-assemble and remove the central tracker.
\item Retract the second half of the inner detector.
\item Separate the ECAL halves from the HCAL end-caps.
\end{enumerate}
The detector would be assembled by simply following the inverse
procedure.

\subsection{Detector Cost Estimate}  


    A rough cost estimate for the TLD is given in table 2.

    The CT cost was taken to be approximately 80\% of the
conservative upper limit (80 MCHF) given in section 8. The ECAL,
HCAL and flux return costs were obtained from the CMS values
scaled by the detector volume and the muon chamber cost was scaled
by area with respect to CMS. The magnet cost was calculated
from a formula provided by Francois Kircher~\cite{kircher} and
the cost of the scintillating tile arrays was taken from
section 9.2. Finally, 5 MDM was added for the luminosity
monitor and 10 MDM for the mechanical structure.

   The total estimated detector cost is approximately 230 MDM.
This is only about half of the cost of the CMS and ATLAS
detectors at the LHC, which reflects the reduced costs for
a smaller, more compact design.

\begin{table}
\centering
\begin{tabular}{|r|r|}
\hline
subdetector  & \hspace{1.0 cm} cost \\
\hline
central tracker & 75 MDM \\
ECAL & 50 MDM \\
HCAL barrel & 30 MDM \\
HCAL end-caps & 4 MDM \\
luminosity monitor & 5 MDM \\
magnet & 10 MDM \\
scint. tile arrays & 3 MDM \\
muon chambers & 23 MDM \\
flux return & 20 MDM \\
mechanical structure & 10 MDM \\
\hline
TOTAL & 230 MDM \\
\hline
\end{tabular} 
\caption{
{\small \bf
A rough cost estimate for the TLD detector. }}
\label{detector cost table}
\end{table}

\subsection{Summary of Example Detector Design}

    A detector design has been presented which attempts to
optimise the choice of subdetector technologies and general
layout for a detector incorporating a compact all-solid state
central tracker. The design seems to achieve the standard of
all-around performance required for exploring physics at
the electron-positron energy frontier. The central tracking
and electromagnetic calorimetry should perform
significantly better than most competing technologies, while
the hadron calorimetry and muon identification performances are,
at mimimum, competitive with rival technologies. The cost of the detector
also seems to be reasonable. Such a detector provides a reasonable
starting place for detailed detector optimization and fine-tuning studies.

\section{Conclusions}

   This report presents an all-solid state central tracker (CT) which
provides both excellent momentum resolution and excellent vertexing.
The tracker would be suitable for the proposed DESY 500 Gev
electron-positron linear collider or a similar accelerator.

   As discussed in section 2, the solid state detectors are arranged
in 10 concentric ellipsoidal shells around the interaction point. The
6 innermost shells consist of pixel detectors and the 4 larger shells
use silicon strip detectors in either small angle stereo or ``U-V-X''
geometry.

   A detailed discussion of the solid state detector elements and
associated electronics was given in section 3. They were found to be
both technologically feasible and suitable for the tracker environment,
with the required specifications representing only a modest
extrapolation from existing detectors.

   The complete absence of gas-based tracking elements allows a very
robust tracker with no time-dependent calibration constants and with a
very fast read-out. The possibility of including even faster trigger
read-out channels means that the position information from the CT can
be made quickly available to the event trigger logic.

   The nearly projective ellipsoidal shell geometry provides smoothly
varying tracking coverage down to small polar angles, avoiding
discontinuities such as barrel-to-endcap cross-over regions.
This geometry is consistent with a pattern recognition algorithm that
is both simple and efficient, and with a light but stiff mechanical structure.
Detailed descriptions of studies for the pattern recognition and
mechanical structure were presented in sections 4 and 5, respectively.

   All the tasks of the central tracker -- track-finding, vertexing and
momentum measurement -- are performed using all the tracking layers
simultaneously, which is both economical and removes many sources
of systematic uncertainty.

   The increased cost of solid state detectors, relative to cheaper
but less performant gas-based detectors, seems to be greatly
compensated by the smaller size of the solid state central tracker.
This allows the CT to be surrounded by smaller, and hence cheaper,
high performance calorimeters and other subdetectors. To illustrate
this, section 9 discusses an example of a reasonably priced yet high
performance collider detector which is based around the all-solid
state CT.

   In summary, the central tracker described in this report seems to
be both affordable and technologically feasible, and would be expected
to provide state-of-the-art tracking performance for the next decade's
collider detectors.

\vskip 0.5 cm
 
\centerline{\large\bf Acknowledgements}
\vskip 0.5 cm

    Many people at the CERN and DESY laboratories have contributed
ideas and expertise to this work. Of these, Jacques Genest deserves
special mention for his invaluable expert assistance with mechanical
simulations of the central tracker. The ECFA/DESY linear collider
workshops have also been
very important in refining and improving the design of the central
tracker. The author would particularly like to thank the CERN and
Hamburg OPAL groups for their help and support.

\pagebreak

\end{document}